\documentclass[conference]{IEEEtran}
\usepackage{cite}
\usepackage[para,online,flushleft]{threeparttable}
\ifCLASSINFOpdf
   \usepackage[pdftex]{graphicx}
 \else
   \usepackage[dvips]{graphicx}
\fi
\usepackage{soul}
\usepackage{amsmath}

\ifCLASSOPTIONcompsoc
 \usepackage[caption=false,font=normalsize,labelfont=sf,textfont=sf]{subfig}
\else
 \usepackage[caption=false,font=footnotesize]{subfig}
\fi

\usepackage{stfloats}
\usepackage{float}

\usepackage{textcomp}  
\usepackage{xcolor}

\begin{document}

\title{A 420 GOPS/W CGRA with a Configurable MAC and Dynamic Truncation}

\author{\IEEEauthorblockN{
*Yi Sheng Chong\IEEEauthorrefmark{2},
*Rakshith Harish\IEEEauthorrefmark{3},
Rajesh Chandrasekhara Panicker\IEEEauthorrefmark{3},
Vishnu P. Nambiar\IEEEauthorrefmark{2},
Anh Tuan Do\IEEEauthorrefmark{2}
}
\IEEEauthorblockA{\IEEEauthorrefmark{2}Institute of Microelectronics, Agency for Science, Technology and Research (A*STAR), Singapore}
\IEEEauthorblockA{\IEEEauthorrefmark{3}Department of Electrical and Computer Engineering, National University of Singapore (NUS), Singapore}
*equally contributed
}

\maketitle

\begin{abstract}
Edge devices demand for highly efficient yet flexible processing capability to handle dynamic real-time workloads. Coarse grain reconfigurable architecture (CGRA) emerges as a suitable accelerator candidate in edge devices, because they are as flexible as general purpose processors and offer high efficiency close to that of domain specific accelerators. However, a typical CGRA requires two cycles for a multiply-and-accumulate (MAC) operation, and workloads such as neural network inference and signal processing involve many MAC operations, resulting in long CGRA processing time. This work proposes a CGRA that has configurable MAC units in the processing elements (PEs) that can perform an addition (ADD) or multiplication (MUL) or a MAC by using the same multiplier and adder, in a single cycle. The readout precision of MAC result can be adjusted by a truncation block. The proposed CGRA is implemented with 40nm CMOS technology. It attains an energy efficiency of 420.6GOPS/W operating at supply of 0.6V and frequency of 21MHz, which is 1.4 times higher than the state-of-the-art.

\end{abstract}

\IEEEpeerreviewmaketitle

\section{Introduction}
Edge computing plays an important role in transforming the way the data generated on the edge devices is stored, processed and transported\cite{IoT_Edge}. Edge devices such as mobile phones and smartwatches are equipped with sensors to deliver user experience. For example, a smartphone runs image recognition to detect objects in a photo captured using a camera, at the same time detects user voice commands recorded using a microphone. To handle various workloads in real time, an edge device requires flexible yet powerful processing capability. This is achieved by the general purpose processors, which attain high performance when scaled according to Moore's law \cite{Moore, EndofMoore}. However, this has led to the 'power wall' phenomenon \cite{PowerWall}, compelling chip designers to prioritize energy efficiency over sheer performance to conserve battery power.

\begin{figure}[t]
    \centering
    \includegraphics[width=0.45\textwidth]{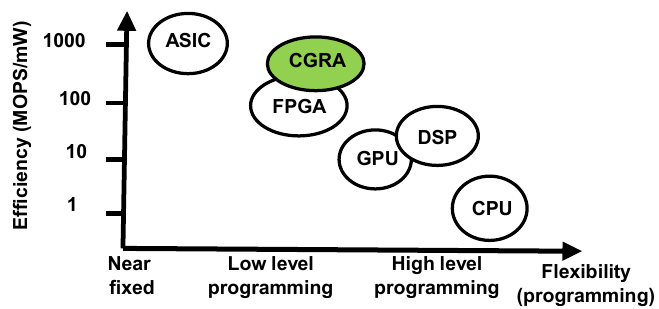}
    \caption{\textcolor{black}{Spectrum of computing architectures where CGRA is relatively flexible and highly programmable \cite{20years}}}
    \label{fig:eff_vs_flexiblity}
    \vspace{-7mm}
\end{figure}

As shown in Fig. \ref{fig:eff_vs_flexiblity}, CGRA offers higher efficiency and flexibility than a general purpose processor and a domain specific accelerator respectively. A typical CGRA consists of a 2D array of processing elements (PEs) that can perform arithmetic, logical and memory operations \cite{CGRASurveyLi2023}. The PEs are connected to their neighbours through reconfigurable switches to send or receive data \cite{PanoramaWijerathne2022}. To run processing on a CGRA, a compiler is employed to generate machine code that includes PE instructions and computation data, to inform the CGRA on the sequence of operations \cite{MorpherWijerathne2022}. The compiler maps any given workload to the CGRA, thus the CGRA is said to be flexible. The CGRA can offer high performance since the compiler can exploit parallelism when mapping a workload, where multiple operations can be executed in parallel in multiple PEs \cite{CGRASurveyLi2023}. Thus, CGRA is a good accelerator candidate in edge devices as it is programmable to process different workloads with high energy efficiency \cite{CGRAReviewAliagha2022,CGRA_Survey}.  




A drawback of current CGRA architectures is that it typically needs two cycles to perform a multiply-and-accumulate (MAC) operation. A MAC operation is typically mapped to two instructions, i.e., one multiplication (MUL) and one addition (ADD) instructions, such that the MUL and ADD are handled by the multiplier and adder in the PE respectively \cite{NPCGRA}. The MAC operations are frequently used in many computations, such as neural network for image recognition \cite{MACinCNNs} and fast Fourier transform (FFT) for digital signal processing \cite{MACinDSP}. A CGRA takes long time to process workloads with high number of MAC operations.


This paper proposes a CGRA PE having a configurable MAC block and a configurable truncation block. The MAC block is configurable to perform either MUL, ADD or MAC operation. The configurable truncation block detects overflow and truncates the accumulation read out to the precision specified by the user. As a result, the CGRA PE reduces the number of cycles when processing workloads that have MAC operations. Our results show improvement in overall CGRA energy efficiency of up to 37\% when running the general matrix multiplication (GeMM) workload.


The rest of this paper is organized as follows. Section \ref{sec:proposed} provides a brief discussion of the proposed CGRA and an in-depth description of the PE, configurable MAC and truncation block in the PE. Section \ref{sec:Perf_eval} presents the performance evaluation methodology and implementation results, followed by the conclusions in Section \ref{sec:conclusion}.

\section{Proposed CGRA}  \label{sec:proposed}

\subsection{Structure of the CGRA}  \label{sec:cgra_struct}

\begin{figure}[t]
    \centering
    \includegraphics[width=0.45\textwidth]{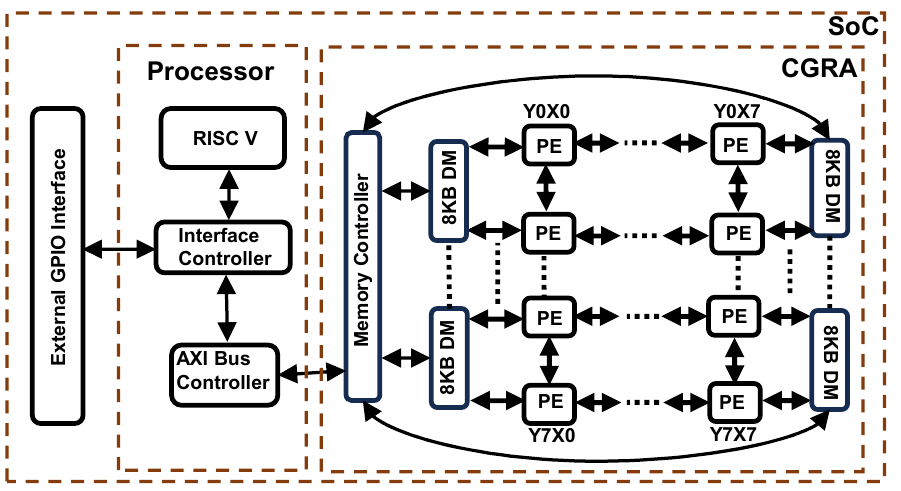}
    \vspace{-3mm}
    \caption{Proposed CGRA that is integrated with a RISC-V based CPU in a system-on-chip}
    \label{fig:CGRA_arch_design}
    \vspace{-5mm}
\end{figure}

Fig. \ref{fig:CGRA_arch_design} presents the proposed CGRA that is based on \cite{HyCubeASSCC2019}. The CGRA is integrated in a system-on-chip where a RISC-V based CPU is used to coordinate the control and communication needs. After a CGRA workload is compiled, the CGRA is updated with a new set of PE instructions and computation data via the the CPU. The proposed CGRA has an 8x8 PE array and data memory (DM) of 64KB. The DM stores the computation data. The PE are meshed connected, where the PE can communicate up to 4 neighbouring PEs. The edge PEs are connected to the DM to receive and send computation data. The PE design is elaborated in Section \ref{sec:pe_structure}.


\subsection{Structure of the PE} \label{sec:pe_structure}

\begin{figure}[t]
    \centering
    \includegraphics[width=0.48\textwidth]{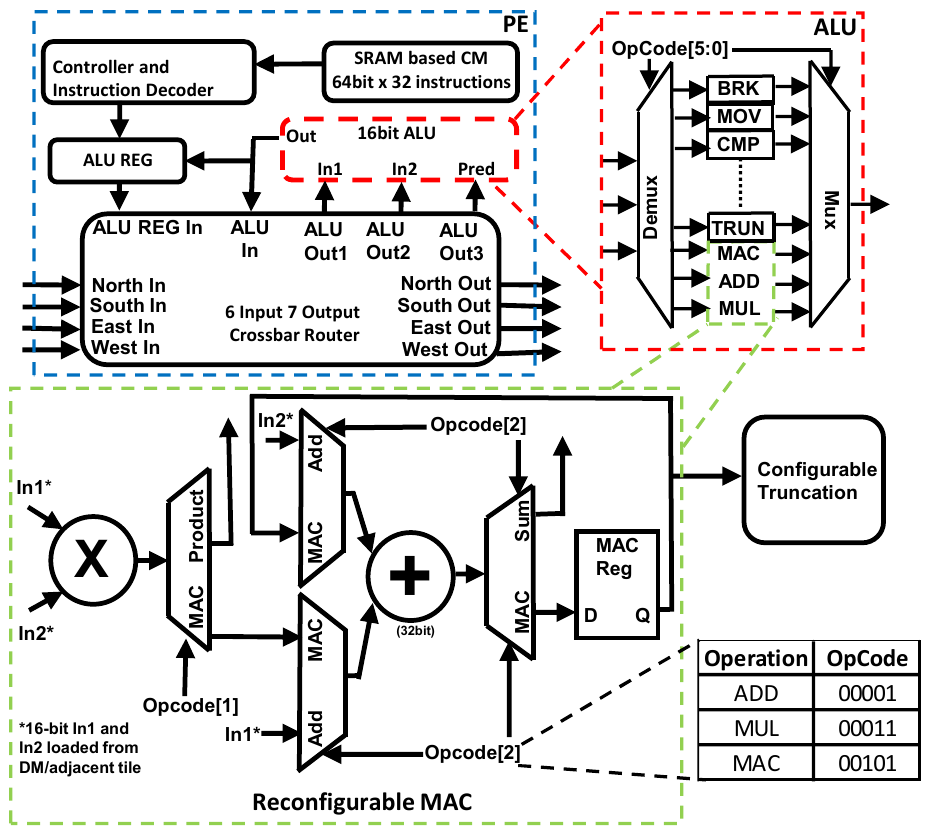}
    \vspace{-3mm}
    \caption{Proposed CGRA PE with the configurable MAC}
    \label{fig:PE_struct}
    \vspace{-6mm}
\end{figure}

Fig. \ref{fig:PE_struct}(a) shows the PE architecture consisting of a configuration memory (CM), arithmetic logic unit (ALU), a control unit and a crossbar router. The control unit fetches and decodes the 64-bit instruction from CM. The instruction configures the ALU operation, and switches in the router. The ALU is capable of performing 18 arithmetic and logical operations such as addition, subtraction, multiplication, logical or and greater-than comparison. The router sends the data coming from neighbouring PEs or the previous ALU results to the ALU for processing. The new ALU results can either be stored temporarily in a register or sent to other PEs depending on the configuration decoded from the instructions. Besides, the router has a buffer-less bypath path that enable data transfer from a PE to a distant PE within one cycle \cite{HyCubeASSCC2019, HycubeKarunaratne2017}. When a PE transfers data to a distant PE, the PEs along the transfer route are configured to enable the bypass path to allow data to directly pass through it. 

In this work, the CGRA is enhanced by introducing a configurable MAC unit with a truncation block, into the PE. The configurable MAC unit works into three modes: MUL, ADD and MAC mode, depending on the opcode decoded from the instructions. The configurable MAC helps to reduce the number of execution cycles when the CGRA processes workloads that have abundant of MAC operations. As discussed earlier, the conventional CGRA takes two instruction cycles to complete a MAC operation as the CGRA PE computes a single operation in a cycle, either MUL or ADD. The configurable MAC is elaborated in Section \ref{sec:configurable_mac}. 



\subsection{Configurable MAC}  \label{sec:configurable_mac}


The configurable MAC, as shown in Fig. \ref{fig:PE_struct}, is a sub-component in the CGRA PE to support three operations: MUL, ADD and MAC, depending on the opcode decoded from the PE instruction. The configurable MAC is designed to reuse the same signed integer multiplier and adder to support all three operations to save hardware resources. Multiplexers (MUXes) are used in the configurable MAC to configure the data path for different operation modes. When performing MUL and ADD, the computation results are sent out from the ALU. When performing MAC operation, the accumulation result is stored in an accumulation register. The bit width of accumulation result is two times larger than the 16-bit input to prevent precision-loss during accumulation. During readout, the accumulation result is passed to the truncation block which is elaborated in Section \ref{sec:trunc_block}, to reduce the accumulation result bit width to match with the data bit width of the CGRA. 

\subsection{Configurable truncation block}  \label{sec:trunc_block}

\textcolor{black}{In the proposed CGRA, the MAC accumulation register in the PE is 32-bit wide as the operand data is 16-bit, to avoid precision loss during accumulation.} Precision truncation reduces the MAC result to 16-bit, so that it is ready to be used by next PE for further processing or store back to the DM. 

\begin{figure}[t]
    \centering
    \includegraphics[width=0.48\textwidth]{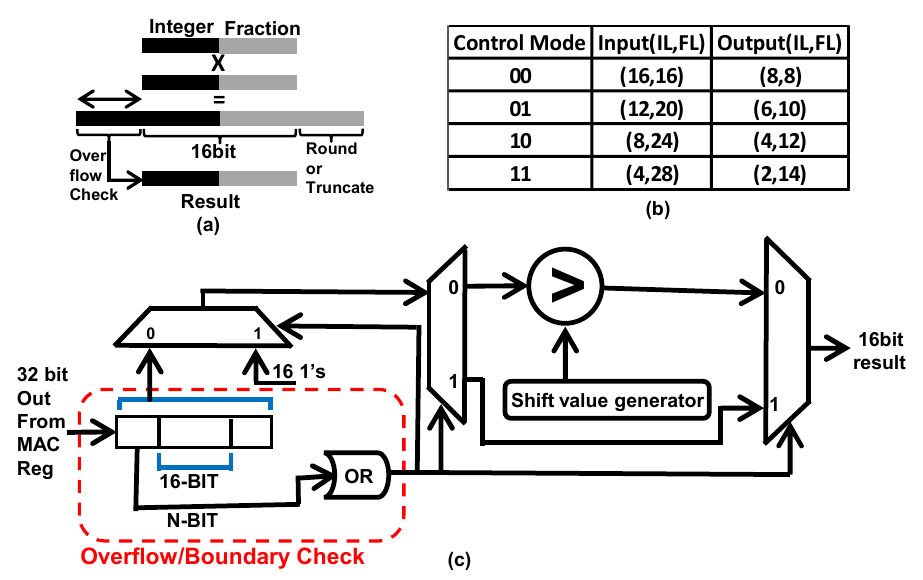}
    \vspace{-3mm}
    \caption{(a) Boundary check mechanism for overflow detection, (b) 4 configuration modes supported by the (c) configurable truncation block.}
    \label{fig:trunc}
    \vspace{-5mm}
\end{figure}

In this work, the truncation is performed after the all accumulation cycles are done. As suggested in \cite{BC_RMAC}, the accuracy when running neural network inference is higher when applying truncation after MAC as compared to truncating after multiplication. Also, in \cite{BC_RMAC}, when handling the overflow situation, the boundary check (BC) mechanism has less impact on the accuracy as compared to the no boundary check (NBC). Hence, the BC mechanism is employed, as illustrated in Fig. \ref{fig:trunc}(a). The BC checks if the overflow bits have one or more ones. If there is, all output bits are set to one, i.e., clipping to maximum value. 


As depicted in in Fig. \ref{fig:trunc}(b), the truncation block supports 4 precision modes configured using 2 bits from the instruction. For instance, in mode 00, to output a fixed point number that is of 8-bit integer and 8-bit fractional (FXP⟨8, 8⟩), the accumulation result is right-shifted by 8 bits and then the lower 16 bits are taken as the outputs. This configurability allows precise control over the truncation and ensures that the appropriate bits are retained according to the desired precision. The other configurations are FXP$\langle 4,12 \rangle$ and FXP$\langle 2,14 \rangle$, since the analysis of CNN accuracy when trained using MNIST dataset and the length of the fractional part for a 16-bit weight in \cite{DNN} shows that using configurations such as FXP$\langle 4,12 \rangle$ and FXP$\langle 2,14 \rangle$ yields satisfactory accuracy results.

\textcolor{black}{Fig. \ref{fig:trunc}(c) shows the design of the configurable truncation block. It first checks if there is any overflow. If yes, the result is set to all ones to represent the maximum clip value. Else, the integer and fractional part are individually truncated. A separate opcode is allocated to truncate MAC results.} 


\section{CGRA Performance Evaluation} \label{sec:Perf_eval}

\subsection{Data flow in conventional and proposed PE}

During runtime, the PE decodes and executes an instruction read from the CM every clock cycle. The PE loops through the CM until it meets the stopping criterion. The PE sets the ALU and router for the desired operation, and to send/receive data as per the decoded configuration. 

\begin{figure}[t]
    \centering
    \includegraphics[width=0.45\textwidth]{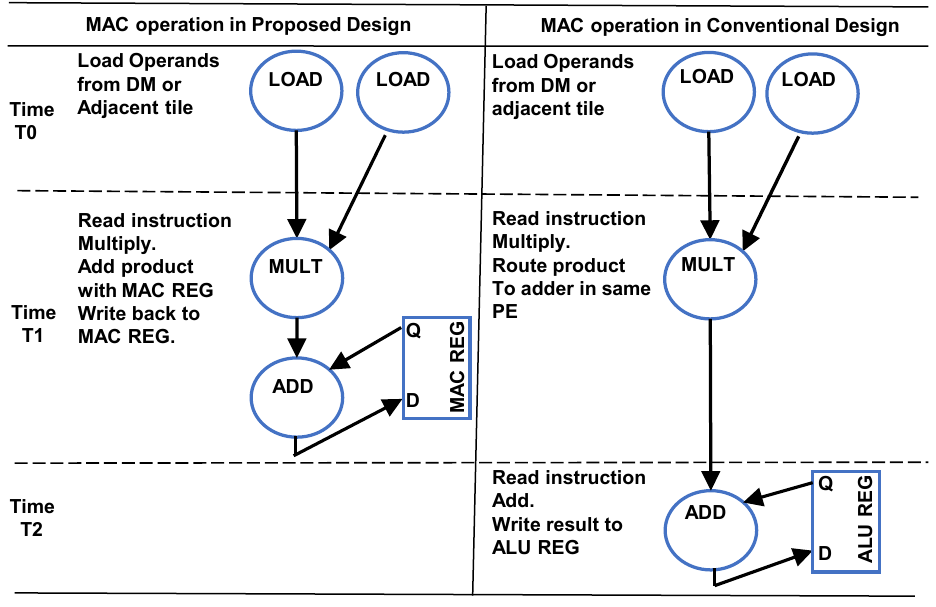}
    \vspace{-3mm}
    \caption{\textcolor{black}{Data flow of the proposed and conventional PE}}
    \label{fig:DFG_compare}
    \vspace{-3mm}
\end{figure}

\textcolor{black}{In a conventional PE, a MAC operation is performed in two cycles. The operation sequence can be represented as a data flow graph as shown in Fig. \ref{fig:DFG_compare}. The input data is loaded from the DM to the ALU input registers. Then, at time step $t_1$, a MUL instruction is executed to perform MUL. The product either stays in the same PE or is transfered to the next PE for accumulation. At time step $t_2$, an ADD instruction is executed to sum the operands taken from the input registers. The addition result is assumed to be stored in a ALU register at the end of $t_2$. Thus, the CGRA requires two time units and two instruction reads to complete a MAC operation.} 


\textcolor{black}{In the proposed PE, a MAC operation is performed in one cycle, as shown in Fig. \ref{fig:DFG_compare}. Activities at time $t_0$ are identical to the conventional PE. Then, the MAC operation occurs at $t_1$ upon the MAC instruction is executed, where MUL is performed on operands, and the product is accumulated with the previous MAC register results, using the adder within the same PE. The new MAC value is then stored in the accumulation register. Hence, by executing just one instruction in a single cycle, the entire MAC operation can be completed within the PE.}

\subsection{Performance of the proposed PE}

\begin{figure}[t]
    \centering
    \includegraphics[width=0.42\textwidth]{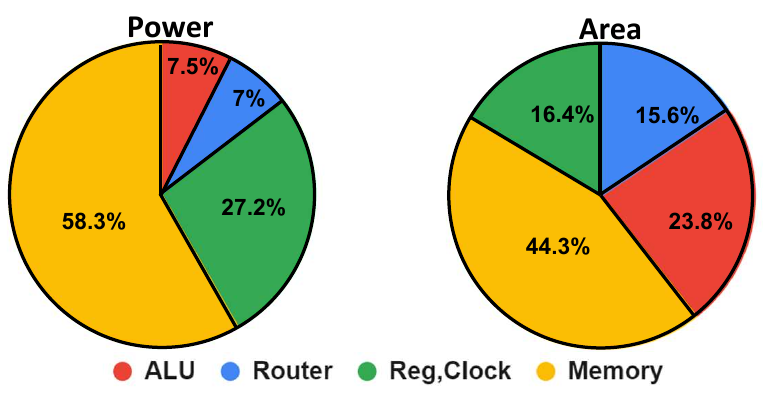}
    \vspace{-3mm}
    \caption{Power and area breakdown of the proposed PE}
    \label{fig:pwr_area_PE}
     \vspace{-3mm}
\end{figure}

\begin{figure}[t!]
    \centering
    \includegraphics[width=0.38\textwidth]{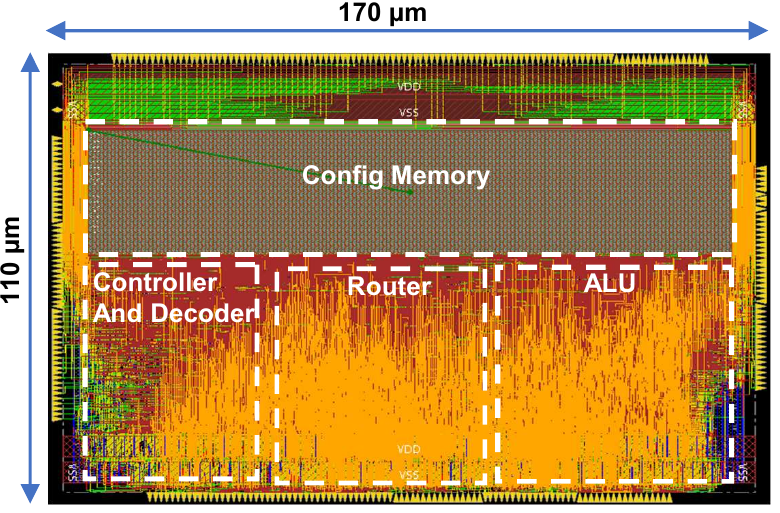}
    \vspace{-3mm}
    \caption{Layout of the proposed PE}
    \label{fig:PE_layout}
    \vspace{-5mm}
\end{figure}

The proposed PE is implemented using Verilog and synthesized using a 40nm technology node. The total area is 0.019 mm\textsuperscript{2}. Fig. \ref{fig:pwr_area_PE} gives the power and area breakdown of the PE. The layout of the PE is shown in Fig. \ref{fig:PE_layout}. \textcolor{black}{Although the proposed PE introduces the reconfigurable MAC to complete MAC operation in one cycle, the hardware path delay of the MAC operation is roughly 3.2ns, which is 1.86 times longer the path delay of the multiplier. Despite the increase in path delay, the proposed CGRA can still meet the target operating frequency of 100MHz in this work.}

The power and energy consumption of the proposed CGRA PE is analyzed when processing only the MAC operations, as compared to the conventional PE. Both conventional and proposed PE are simulated with the GeMM workload and their power consumption are obtained using Cadence Joules. For instance, in an 8x8 GeMM workload, the average power consumption of the proposed CGRA PE is 0.74mW when processing 512 MAC operations, which is similar to the power consumption of the conventional PE. However, the energy consumption of the proposed PE is 0.34 times smaller than the conventional PE, when processing the same number of MAC operations for 8x8 GeMM. As shown in Fig. \ref{fig:DFG_compare}, the proposed PE only requires one memory access for one MAC operation, yet, the conventional PE requires 2 memory accesses to obtain the instructions to complete one MAC operation. Based on our simulation, the energy consumption of CM read access is 5.14 and 2.19 times than that of ADD and MUL respectively. Thus, the energy savings of the proposed PE is 37\%.

\subsection{Performance of the CGRA}

To evaluate the performance impact of the proposed PE on the CGRA performance, the CGRA is simulated using the GeMM workload. GeMM serves as a foundational computation in numerous applications such as speech processing and fast Fourier transform \cite{FFTonGEMMPedram2013}. It is a well-known fundamental operation in linear algebra and is commonly employed as a benchmark to assess the performance of high-performance computing systems \cite{FFTonGEMMPedram2013}. It holds a significant position in the 13-dwarfs categorization of computational patterns \cite{NPCGRA}. The equation of GeMM is given in Equation \ref{eqn:gemm}. The proposed CGRA utilizes the compiler proposed in \cite{MorpherWijerathne2022} to compile the GeMM workload.
\begin{equation}  \label{eqn:gemm}
   A[\ ] = B[\ ] \times C[\ ]
\end{equation}
where $A[\ ]$ is the resultant matrix, $B[\ ]$ and $C[\ ]$ are the input matrices. 

\begin{table*}[t]
\centering
\caption{Comparison with state-of-the-art CGRA}
\begin{threeparttable}
\begin{tabular}{|c|c|c|c|c|c|c|c|c|}
\hline
& Amber\cite{Amber} & SSCL'20\cite{2D} & ISSCC'19\cite{2.2} & TVLSI'18\cite{SDT} & ASSCC'19\cite{HyCubeASSCC2019} & JSSC'20\cite{MIMO} & \textbf{Proposed} \\
\hline
Tech (nm) & 16 & 28 & 22 & 55 & 40 & 28 & \textbf{40} \\ 
\hline
Voltage (V) & \textcolor{black}{0.84} & 0.6  & 0.8 & NA & 1.1 & 0.9 & \textbf{1}  \\ 
\hline
Frequency (MHz) & 955 & 89 & 36 & 450 & 853 & 800 & \textbf{100} \\ 
\hline
PE count & 384 & 120 & 15 & 30 & 16 & 64 & \textbf{64} \\ 
\hline
Throughput (GOPS)  & 367 & 14.1 & 145 & 77.4  & 6.5   & 0.9 & \textbf{9.5} \\ 
\hline
Power (mW) & NA & 45.9 & NA & 1526  & 6.5 & 537 & \textbf{47.6} \\ 
\hline
Efficiency (GOPS/W) & \textcolor{black}{538@0.84V} & 307@0.6V & 978@0.48V & 50.8  & 90 & 196  & \textbf{\begin{tabular}[c]{@{}c@{}}199.2@1V\\420.7@0.6V\end{tabular}} \\ 
\hline
Norm. Efficiency (GOPS/W) \tnote{1} & 86 & 150.4 & 295.8 & 96 & 90 & 96 & \textbf{420.7} \\
\hline
\end{tabular}
\begin{tablenotes}
\item [1] Norm. Efficiency = Efficiency $\times (\dfrac{\text{node}}{40\text{nm}})^2$ \textcolor{black}{, derived using general scaling in Table 1 from \cite{ScalingStillmaker2017} for coarse performance estimation}
\end{tablenotes}
\end{threeparttable}
\label{tab:comparison}
\vspace{-5mm}
\end{table*}

\begin{figure}[t]
    \centering
    \includegraphics[width=0.48\textwidth]{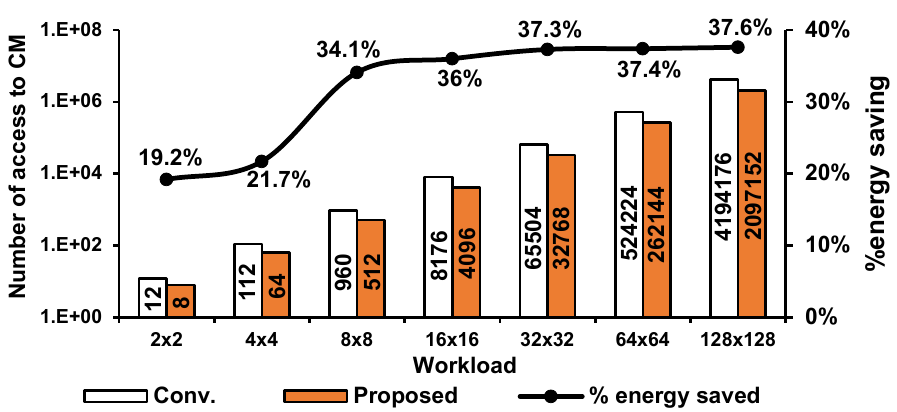}
    \vspace{-3mm}
    \caption{Number of instruction reads and energy savings when the proposed CGRA runs various GeMM workloads as compared to conventional CGRA}
    \label{fig:CM_access_saved}
     \vspace{-5mm}
\end{figure}


Fig. \ref{fig:CM_access_saved} shows the energy saved by the proposed CGRA when running GeMM workload. The energy savings are mainly contributed by the reduction in instructions read from the CM in the PEs. Considering the 8x8 GeMM workload as an example, the conventional CGRA requires 512 MUL and 448 ADD, totaling to 960 arithmetic instruction reads from the CM. For the proposed CGRA, this number reduces to only 512 MACs, which is a 54\% drop. However, the total instructions generated by the compiler are more than 512, due to the overhead of control instructions for address generation for input and output data access from the DM. This effectively results in one MAC performed in 1.35 clock cycles. Overall, the throughput achieved by the proposed CGRA is 9.5GOPS, consuming an average power of 47.6mW at 1V supply. The peak efficiency observed is 420.7GOPS/W at 0.6V and 21MHz.

Table \ref{tab:comparison} compares the proposed CGRA with the state-of-the-art. The efficiency is normalized to 40nm technology node for fair comparison. The proposed CGRA attains peak efficiency of 420.7 GOPS/W, operating at 0.6V and 21MHz, which is about 1.4 times higher than \cite{2.2}. The design from \cite{2.2} comes closest to this work, achieving high energy efficiency due to distributed memories at the lowest hierarchy level, high PE count and multi-directional routing. 

\section{Conclusion} \label{sec:conclusion}

A CGRA that employs PEs using the proposed configurable MAC with truncation block has been presented. The configurable MAC block allows the CGRA to complete a MAC operation in one cycle, reducing the number of processing cycles, whilst reusing the same hardware for MUL and ADD. The configurable truncation block is included to allow users to choose the precision when the MAC result accumulated at full precision leaves the PE. The proposed CGRA, synthesized using a 40nm technology node, attains efficiency of 420.6GOPS/W operating at supply of 0.6V and frequency of 21MHz, which is 1.4 times higher than the state-of-the-art.


\section*{Acknowledgement}
This research is supported by the National Research Foundation, Singapore (CRP23-2019-0003).

\bibliographystyle{IEEEtran}
\bibliography{references}

\end{document}